\newcommand{\beq}{\begin{equation}}
\newcommand{\eeq}{\end{equation}}
\newcommand{\bea}{\begin{eqnarray}}
\newcommand{\eea}{\end{eqnarray}}

\newcommand{\gsim}{\lower.7ex\hbox{$\;\stackrel{\textstyle>}{\sim}\;$}}
\newcommand{\lsim}{\lower.7ex\hbox{$\;\stackrel{\textstyle<}{\sim}\;$}}

\documentclass[aps,prev,twocolumn,preprintnumbers,floatfix,nofootinbib]{revtex4-1}
\usepackage{graphicx}
\usepackage{epstopdf}
\usepackage{mathrsfs}
\usepackage{amssymb}
\usepackage{verbatim}
\usepackage{color}
\usepackage{multirow}
\usepackage{axodraw}

\def\stacksymbols #1#2#3#4{\def\theguybelow{#2}
    \def\vp{\lower#3pt}
    \def\sp{\baselineskip0pt\lineskip#4pt}
    \mathrel{\mathpalette\intermediary#1}}

\def\intermediary#1#2{\vp\vbox{\sp
     \everycr={}\tabskip0pt
     \halign{$\mathsurround0pt#1\hfil##\hfil$\crcr#2\crcr
              \theguybelow\crcr}}}

\def\be{\begin{equation}}
\def\ee{\end{equation}}
\def\bea{\begin{eqnarray}}
\def\eea{\end{eqnarray}}

\def\sp{\;\;\;,\;\;\;}

\def\lsim{\raise0.3ex\hbox{$\;<$\kern-0.75em\raise-1.1ex\hbox{$\sim\;$}}}
\def\gsim{\raise0.3ex\hbox{$\;>$\kern-0.75em\raise-1.1ex\hbox{$\sim\;$}}}

\def\inbar{\,\vrule height1.5ex width.4pt depth0pt}

\def\IC{\relax\hbox{$\inbar\kern-.3em{\rm C}$}}
\def\IQ{\relax\hbox{$\inbar\kern-.3em{\rm Q}$}}
\def\IR{\relax{\rm I\kern-.18em R}}
 \font\cmss=cmss10 \font\cmsss=cmss10 at 7pt
\def\IZ{\relax\ifmmode\mathchoice
 {\hbox{\cmss Z\kern-.4em Z}}{\hbox{\cmss Z\kern-.4em Z}}
 {\lower.9pt\hbox{\cmsss Z\kern-.4em Z}}
 {\lower1.2pt\hbox{\cmsss Z\kern-.4em Z}}\else{\cmss Z\kern-.4em Z}\fi}

\def\comment#1{}
\def\to{\rightarrow}

\def\u1x{U(1)_X}
\newcommand{\nc}{\newcommand}
\nc{\LL}{L}
\nc{\vv}{\tilde{v}}
\nc{\ccdot}{\!\cdot\!}
\nc{\gsm}{G_{SM}}
\nc{\vfive}{\mathbf{5}\oplus\mathbf{\overline{5}}}
\nc{\vten}{\mathbf{10}\oplus\mathbf{\overline{10}}}
\nc{\zhol}{Z^{\rm hol}}
\nc{\xfb}{\,{\rm fb}}

\begin{document}

%
%

\preprint{KCL-PH-TH/2013-33,LCTS/2013-22,CERN-PH-TH/2013-241,ACT-8-13,MIFPA-13-46,UMN--TH--3309/13,FTPI--MINN--13/38}

\vspace*{1mm}

\title{A No-Scale Framework for Sub-Planckian Physics}

\author{John~Ellis$^{a}$}
\email{John.Ellis@cern.ch}
\author{Dimitri~V.~Nanopoulos$^{b}$}
\email{dimitri@physics.tamu.edu}
\author{Keith A. Olive$^{c}$}
\email{olive@physics.umn.edu}

\vspace{0.1cm}
\affiliation{
${}^a$ Theoretical Particle Physics and Cosmology Group, Department of
  Physics, King's~College~London, London WC2R 2LS, United Kingdom;\\
Theory Division, CERN, CH-1211 Geneva 23,
  Switzerland
 }
 \affiliation{
${}^b$ 
 George P. and Cynthia W. Mitchell Institute for Fundamental Physics and Astronomy,
Texas A\&M University, College Station, TX 77843, USA;\\
Astroparticle Physics Group, Houston Advanced Research Center (HARC), Mitchell Campus, Woodlands, TX 77381, USA;\\
Academy of Athens, Division of Natural Sciences,
28 Panepistimiou Avenue, Athens 10679, Greece,}
 
\affiliation{
${}^c$ 
William I. Fine Theoretical Physics Institute, School of Physics and Astronomy,\\
University of Minnesota, Minneapolis, MN 55455,\,USA}

\begin{abstract} 

We propose a minimal model framework for physics below the Planck scale with the following features:
$(i)$ it is based on no-scale supergravity, as favoured in many string compactifications,
 $(ii)$ it incorporates Starobinsky-like inflation, and hence is compatible with constraints from the Planck satellite,
 $(iii)$ the inflaton may be identified with a singlet field in a see-saw model
 for neutrino masses, providing an efficient scenario for
 reheating and leptogenesis, $(iv)$ supersymmetry breaking occurs with an arbitrary scale and a
 cosmological constant that vanishes before radiative corrections, $(v)$ regions of the model parameter space 
 are compatible with all LHC, Higgs and dark matter constraints. 
 \end{abstract}

\maketitle


\setcounter{equation}{0}




There are two extensions of the Standard Model of particle physics that seem to us particularly
well motivated: heavy singlet (right-handed) neutrinos to generate mixing and small masses for the
standard left-handed neutrinos via the see-saw mechanism~\cite{seesaw}, and supersymmetry~\cite{SUSY} to stabilize
the electroweak scale, facilitate grand unification, etc.. Likewise, there are three well-motivated 
extensions of the standard Big Bang model of cosmology: baryogenesis~\cite{Sakharov}, dark matter~\cite{darkmatter}, 
and inflation~\cite{inflation} to explain the large-scale homogeneity of the universe and the fluctuations in the cosmic
microwave background (CMB). There are also popular connections
between these extensions of the Standard Model: for example, CP violation in the decays of the
heavy singlet (right-handed) neutrinos might have generated a lepton asymmetry that would
subsequently have been converted into a baryon asymmetry via sphaleron interactions \cite{FY},
supersymmetry provides a natural dark matter candidate \cite{gold,ehnos}, and the inflaton might have been
a heavy singlet sneutrino \cite{ERY}.

The purpose of this paper is to propose a (the?) minimal framework that combines all these
desirable features. Low-energy particle physics is described by the minimal ${\cal N} = 1$ supersymmetric
extension of the Standard Model (MSSM), which is supplemented by a supersymmetric
Type-1 see-saw heavy neutrino sector. One of the heavy sneutrinos is identified as the
inflaton~\cite{ERY}, and its decays generate a lepton (and hence baryon) asymmetry~\cite{FY}. The
inflaton is embedded in a no-scale supergravity sector~\cite{EENOS} together with another
singlet field~\cite{ENO6,ENO7,others}, in a form that could emerge naturally as the low-energy effective field theory
obtained from some string compactification~\cite{Witten}. This form yields an effective potential and
hence a spectrum of perturbations identical with the Starobinsky $R + R^2$ model~\cite{R2}, which
is comfortably compatible with data from the Planck satellite~\cite{Planck}.
We display a choice of superpotential for these fields that
generates supersymmetry breaking with an arbitrary magnitude and zero vacuum energy 
at the tree level. We show that there are regions of the model parameter space that are
compatible with the measured mass of the Higgs boson, supersymmetric dark matter
and the absence of a supersymmetric signal at the LHC.

Since we incorporate supersymmetry and consider cosmology, our model is 
formulated within an ${\cal N} = 1$ supergravity framework~\cite{sugra}. This is characterized by a Hermitian K\"ahler
potential $K(\phi_i, \phi^*_j)$ and a holomorphic superpotential $W(\phi_i)$, where the $\phi_i$
are complex scalars fields in chiral supermultiplets, in the
combination $G \equiv K + \ln W + \ln W^*$. In order to have an effective potential
that vanishes over a continuous range of parameters in the presence of supersymmetry
breaking, and motivated by generic string compactifications~\cite{Witten}, we choose a no-scale
form for $K$ that has an underlying SU($N, 1$)/SU($N$)$\times$U($1$) symmetry~\cite{noscale}. For the
purposes of our discussion here, we display an example with $N = 2$ that can be
extended to include, e.g., the fields of the MSSM. In this case there are two
equivalent field representations that are frequently used: 
\beq
K = - 3 \ln (T + T^* -  |\phi|^2/3 + \dots )
\eeq
and 
\beq
K = - 3 \ln (1 - |y_1|^2/3 -  |y_2|^2/3 - \dots ). 
\eeq
In the first formulation, the $T$ field
may represent a string modulus and $\phi$ a generic matter field, whereas in the second
formulation both $y_{1,2}$ may be matter fields.

It was shown in~\cite{ENO6,ENO7} that the $N = 2$ no-scale model can yield an effective
potential identical with that of the Starobinsky $R + R^2$ model for various suitable choices
of the superpotential $W$~\cite{Cecotti}. Here we discuss one example that was introduced in~\cite{ENO6}
and explored in more detail in~\cite{ENO7}, namely
\begin{equation}
W \; = \; M \left[ \frac{y_1^2}{2} \left(1+\frac{y_2}{\sqrt{3}} \right) - \frac{y_1^3}{3 \sqrt{3}}  \right]\, .
\label{W1}
\end{equation}
Unless explicitly denoted, we will work in Planck units $M_P^2 = 1$, where $M_P^{-2} = 8\pi G_N$ refers to the normalized Planck mass.
Eq.\ \ref{W1} represents the minimal Wess-Zumino (WZ) model~\cite{SUSY} for the no-scale field $y_1$, 
with a minimal interaction $y_1^2 y_2$ with the other no-scale field $y_2$.
Note that the modular factor $(1 + y_2/\sqrt{3})$ arises from the field redefinition from
the $(T,\phi)$ basis.  In general the modular weight of a superpotential term of field dimension
$D$ is $3-D$. 

The model described by Eq.\ \ref{W1} reproduces the effective potential of the Starobinsky model for
${\cal R}e~y_1$ when a suitable stabilizing term $\propto y_2^4$ is introduced into the
no-scale K\"ahler potential~\cite{EKN3,ENO7,kl}:
\beq
K \; = \; - 3 \ln \left(1 - \frac{|y_1|^2 + |y_2|^2}{3} +  \frac{ |y_2|^4}{\Lambda^2} \right) \, ,
\label{K1g}
\eeq
with $\Lambda \lsim 0.1 M_P$, as discussed in~\cite{ENO7}.

The no-scale structure explored here is subject to radiative corrections,
which depend on the ultraviolet completion of the theory,
and are thought to play an essential r\^ole in determining the scale of supersymmetry breaking~\cite{ELNT}.
Since discussion of the ultraviolet completion lies beyond the scope of this paper, which is concerned with
sub-Planckian physics, and since the scale of inflation is much higher than the supersymmetry-breaking scale, 
we leave a more complete discussion of the structure of these radiative corrections for future work.

Having discussed how inflation may be embedded in a no-scale framework typical of a generic
string compactification, we now discuss how it may be connected with lower-energy physics.
The first issue is the possible identification of the inflaton field $y_1$ with some field
appearing in (a motivated extension of) the Standard Model. 

\begin{figure}[t]
\begin{center}
\begin{picture}(200,120)(0,0)
\SetWidth{1.0}
\Line(25,75)(75,75)
\Text(75,82)[]{$g$}
\Line(75,75)(125,100)
\Text(125,90)[]{$\nu$}
\DashLine(75,75)(125,50){4}
\Text(85,60)[]{${\tilde \nu}$}
\Text(110,49)[]{${\tilde N}$}
\Text(50,90)[]{$\chi$}
\Text(100,63)[]{$\times$}
\Text(110,70)[]{$\lambda v M$}
\Line(125,50)(175,75)
\Text(125,43)[]{$\uparrow$}
\Text(113,34)[]{$M/M_P$}
\Text(135,63)[]{${N}$}
\Text(175,65)[]{$\nu$}
\Text(150,63)[]{$\times$}
\Text(150,54)[]{$\uparrow$}
\Text(150,46)[]{$\downarrow$}
\Text(160,50)[]{$\lambda v$}
\Line(125,50)(175,25)
\Text(135,37)[]{${N}$}
\Text(150,38)[]{$\times$}
\Text(175,35)[]{$\nu$}
\end{picture} 
\end{center}
\vspace{-1cm}
\noindent
\caption{\it Diagram for $\chi \to 3 \nu$ decay. The $\times$ symbols
denote light $\nu$ - heavy $N$ mixing via a Dirac Yukawa coupling
as expected in a Type-I see-saw model.}
\label{fig:3nu}
\end{figure}
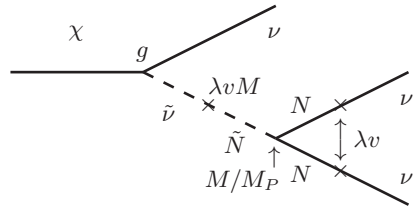

The forms of the interactions
in (\ref{W1}) require $y_1$ to be a gauge-singlet field, and the minimal possibility is to
identify it with one of the singlet (right-handed) neutrinos in the minimal Type-1 see-saw
extension of the Standard Model (a singlet sneutrino). 
Thus we can add to the superpotential of the standard model a Yukawa-like term
\beq
W_Y = \lambda H_2^i L_i y_1
\label{Yuk}
\eeq
where $H_2$ is the same Higgs doublet giving mass to the up-like quarks, and $L$ is 
a left-handed lepton doublet. The index $i$ is an SU(2) index. In principle, we expect
3 right right-handed singlet neutrinos, but we associate only one (or one linear combination
of the three) with the inflaton.
Remarkably, the order of magnitude of the mass
parameter $M$ in (\ref{W1}) required to yield the right size of primordial scalar
perturbations, namely $\sim 10^{13}$~GeV, lies within the range favoured in such
see-saw models of neutrino masses and mixing. Furthermore,
previous studies have shown that such a sneutrino inflationary scenario can comfortably
generate a lepton asymmetry of the size required for successful leptogenesis~\cite{ERY,ant}.

However, there is one issue that needs to be addressed, namely that of lepton-number
violation. When $y_1$ is identified as a sneutrino, in addition to the Majorana mass-generating
$\Delta L = 2$ interaction that conserves $R$ parity, there is a trilinear $\Delta L = 3$
interaction that violates $R$ parity, implying that the LSP is in principle unstable.
However, its lifetime would be very long~\cite{CEM}. If we identify the LSP with the
lightest neutralino $\chi$, on the basis of diagrams such as that in Fig.~\ref{fig:3nu}
we estimate that its lifetime is
\beq
\tau_{LSP} \; \sim \; \frac{192 \pi^3 M_P^2 (M {\tilde m})^{4}}{g^2 \lambda^6 m_\chi^5 v^6} \, ,
\label{tauLSP}
\eeq
where $v$ is the Higgs vacuum expectation value (vev), ${\tilde m}$ is a typical 
supersymmetry breaking soft mass, $\lambda$ is the Yukawa coupling between the light doublet and heavy singlet neutrinos
and $g$ is a gauge coupling. Taking $M \sim 10^{13}$~GeV, ${\tilde m} \sim 1$~TeV and
$\lambda = {\cal O}(1)$, this estimate yields an LSP lifetime exceeding $10^{40}$~years.
For smaller $\lambda$, as required to obtain a suitable reheat temperature, 
the lifetime of the neutralino will be substantially longer.

As the inflaton is directly coupled to Standard Model fields through (\ref{Yuk}),
reheating and leptogenesis can proceed through the out-of-equilibrium decay
of the heavy singlets~\cite{FY}. The decay rate of $y_1$ can be approximated
as $\Gamma_D \simeq \lambda^2 M/(16 \pi)$, which is smaller than the 
Hubble rate at the end of inflation, $H_I \simeq M/2$. The (instantaneous) 
reheating temperature, $T_R$ is easily estimated to be
\beq
T_R \simeq \left( \frac{5}{\pi^3 N} \right)^{1/4} \left(\frac{\lambda^2 M M_P}{\sqrt{32 \pi}} \right)^{1/2} \sim 2 \times 10^{14} \lambda 
 {\rm GeV} \, ,
 \eeq 
 assuming decay at $\Gamma_D \simeq \frac{3}{2} H$ where $N$ is the number of 
 degrees of freedom at the time of decay, which we take to be the MSSM value of 885/4.
 
 High reheat temperatures are known to give rise to cosmological problems associated
 with gravitino production.
Although gravitinos never dominate the energy density of the Universe if $m_{3/2} \sim 10$~TeV, 
gravitino decay to neutralinos, $\chi$,  can lead to an overabundance of the lightest supersymmetric particle
that provides cold dark matter~\cite{ehnos}. The abundance of gravitinos can be determined in terms of the 
reheat temperature as~\cite{ehnos,thermal}
\beq
\frac{n_{3/2}}{s} \simeq 2.4 \times 10^{-12} \left( \frac{T_R}{10^{10} {\rm GeV}} \right) \, ,
\eeq
where $s$ is the entropy density and we have assumed that the gravitino is much heavier than the gluino. 
In order to satisfy the upper limit on the abundance of neutralinos: $\Omega_\chi h^2 < 0.12$, we must ensure that~\cite{ego}
\beq 
\frac{n_{3/2}}{s} \lesssim 4.4 \times 10^{-13} \left( \frac{1 {\rm TeV}}{m_\chi} \right) \, ,
\eeq
which leads to an upper limit on the Yukawa coupling $\lambda$:
\beq
\lambda \lesssim 10^{-5} \, .
\eeq
This result implies that one of neutrino Yukawa couplings may be comparable
to the electron Yukawa coupling and suggests that the mass of lightest neutrino 
generated by the seesaw mechanism should be of order $10^{-10} - 10^{-9}$~eV. 
We note that the small value of $\lambda$ assures out-of equilibrium decay
and hence the possibility of successful leptogenesis.

Sneutrino inflation is not the only possibility within our no-scale framework.
In some extensions of the Standard Model with extended gauge groups the
heavy neutrinos are not singlets, but there may be other gauge-singlet
superfields. One such example is the minimal flipped SU(5) GUT~\cite{F5}, where the
heavy neutrinos are embedded in $\mathbf{10}$ representations $F_i$, and the inflaton may
be identified with one of the other gauge-singlet fields $\phi_i$ in the model, which
are expected to have trilinear couplings. The relevant couplings are of the form
$\lambda_{6ij} F_i {\bar H} \phi_j + \lambda_{8ijk} \phi_i \phi_j \phi_k$, where
${\bar H}$ denotes a $\mathbf{\overline {10}}$ Higgs representation with a GUT-scale vev, $V$,
and one of the $\phi$ has a vev of similar magnitude. In this case, there is mixing
between the heavy neutrinos and the singlet fields~\cite{ENO2}, the lightest of their mass eigenstates
may well have a mass $\sim 10^{13}$~GeV as required for the inflaton, 
and it would in general have a trilinear self-coupling as required (\ref{W1}) in our
no-scale Starobinsky inflationary scenario. Previous studies have
shown that this model can also lead to successful leptogenesis~\cite{ENO2},
but we do not enter here into details of the heavy sneutrino/singlet sector.

We now address another issue concerning lower-energy physics, namely
supersymmetry breaking. One of the attractive features of the original no-scale model~\cite{noscale}
was that it accommodated local supersymmetry breaking in the form of an arbitrary
gravitino mass $m_{3/2} = e^{G/2}$ with zero vacuum energy (cosmological constant)
before the calculation of quantum corrections. This even suggested a perturbative mechanism
for the dynamical determination of $m_{3/2}$~\cite{ELNT}. Supersymmetry breaking can be
accommodated within our framework by simply adding to the superpotential (\ref{W1}) a constant (modular weight 3)  term
\begin{equation}
W_{SSB} \; = \; r \left( 1 + \frac{y_2}{\sqrt{3}}\right)^3 \, .
\label{WSSB}
\end{equation}
It is easy to check that, as $\Lambda \to \infty$ in the K\"ahler potential (\ref{K1g}) the parameter $r$
does not appear in the effective potential, implying that this is a further generalization of
the no-scale Starobinsky models discussed previously in~\cite{ENO6,ENO7}. When
$\Lambda$ is finite, the effective potential does depend on $r$, but this dependence drops out
when $y_2 = 0$, as we enforce by requiring $\Lambda \lsim 0.1 M_P$~\cite{ENO7}.
It is also easy to check that the magnitude of local supersymmetry breaking: $m_{3/2} = r$
while the vacuum energy vanishes, as in the original no-scale model~\cite{noscale}. In (\ref{WSSB})
we treated $r$ as a parameter to be derived (presumably) from string dynamics.
It would be interesting to resurrect the possibility of determining $r = m_{3/2}$ dynamically
via quantum corrections in the low-energy effective theory, but we do not discuss that
possibility here. Note that, in the presence of both non-vanishing $r$ and finite $\Lambda$,
the modulus $y_2$ is strongly stabilized with a mass (for both real components)
$m_{y_2} = 6\sqrt{2} r/\Lambda = 6\sqrt{2} m_{3/2} M_P/ \Lambda \gg m_{3/2}$. 

We consider next the extension of the model (\ref{W1}, \ref{K1g}, \ref{WSSB})
to Standard Model particles and their supersymmetric partners $y_{SM}$. In the spirit
of the no-scale approach we embed them within the logarithm in the K\"ahler potential:
\begin{equation}
K \; \ni \; - 3 \ln ( 1 - |y_1|^2/3 - |y_2|^2/3 - |y_{SM}|^2/3 - \dots) \, .
\label{nsKahler}
\end{equation}
We assume that the superpotential for the
Standard Model superfields has the form
\begin{equation}
W_{SM} \; = \; W_2(y_{SM}) \left( 1 + \frac{y_2}{\sqrt{3}}\right)^\beta + W_3(y_{SM}) \left( 1 + \frac{y_2}{\sqrt{3}}\right)^\alpha \, ,
\label{WSM}
\end{equation}
where $W_{2,3} (y_{SM})$ are bi/trilinear in the Standard Model superfields, respectively, in which case the
soft supersymmetry-breaking mass-squared, bilinear and trilinear parameters are
\begin{equation}
m_0 \; = \; 0, \; B_0 \; = \; (\beta - 1) m_{3/2}, \; A_0 \; = \; \alpha m_{3/2} \, .
\label{mBA}
\end{equation}
The choice of modular weights $\alpha = 0$, $\beta = 1$ for the tri-linear and bilinear terms respectively,  corresponds to the pure no-scale option $m_0 = B_0 = A_0 = 0$, in which case
the only possibility for introducing supersymmetry-breaking into the low-energy sector
is via a non-minimal gauge kinetic term generating non-zero gaugino masses $m_{1/2} \ne 0$.

We consider the generic option that $m_{3/2} \gg m_{1/2}$, in which case there are no Big Bang
nucleosynthesis constraints on gravitino decays, and we assume that the LSP is the
lightest neutralino, so there are no problems associated with the large reheating 
temperature discussed earlier.  In this case, however, the no-scale choices $\alpha=0$
and $\beta = 1$ are the only phenomenologically viable choices.
For $A_0 \propto m_{3/2}$ (i.e., $\alpha \ne 0$) and  $m_{3/2} \gg m_{1/2}$, renormalization 
group evolution inevitably leads to tachyonic sfermion masses.
Similarly, for $\beta \ne 1$, there is no value of $\tan \beta$ which minimizes the Higgs potential.

It was shown in~\cite{EMO2} that such no-scale boundary conditions
are compatible with low-energy constraints on supersymmetry if there is a grand
unification group at high scales that is broken down to SU(3)$\times$SU(2)$\times$U(1)
at an appropriate high-energy scale. There are three relevant parameters in such a
scenario based on SU(5): the scale $M_{in}$ at which the no-scale boundary conditions are applied,
and two SU(5) Higgs superpotential parameters $\lambda, \lambda^\prime$:
$W \ni \lambda {H} {\Sigma} {\bar H} + (\lambda^\prime/6) {\rm Tr}{\Sigma^3}$, where
$H, {\bar H}$ and $\Sigma$ are $\mathbf{5}, \mathbf{\bar 5}$ and $\mathbf{24}$
Higgs representations, respectively. Regions
of the no-scale SU(5) parameter space that were compatible with accelerator
constraints on supersymmetric particles and the cosmological LSP density were
studied in~\cite{EMO2}. Since then, the LHC has imposed additional constraints, in
particular via the Higgs discovery~\cite{Higgs} and the decay $B_s \to \mu^+ \mu^-$ \cite{bmm}. 

\begin{figure}[h!]
\includegraphics[scale=0.45]{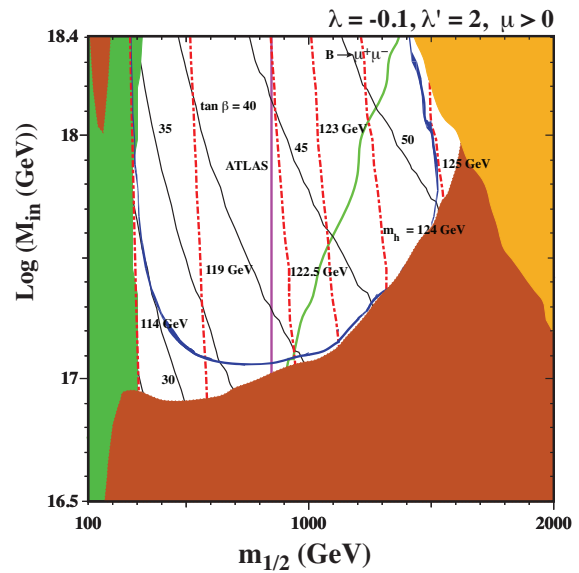}
\caption{\it The $(m_{1/2}, M_{in})$ plane in the SU(5) no-scale
model with $m_0 = A_0 = B_0=0$ at $M_{in}$, for $\lambda^\prime = 2$ and $ \lambda = - 0.1$.
The brown regions are excluded because the LSP is charged,
the green regions are excluded by $b \to s \gamma$, the renormalization-group equations are
unstable in the ochre region. The relic cold dark matter density lies within the cosmological range
along the (dark blue) strips,  the continuous (purple) line is the ATLAS 95\% CL limit on supersymmetric 
particles, the dash-dotted (red) lines are contours of $m_H$ as calculated using {\tt FeynHiggs}~\cite{FH},
and the solid (green) line marks the 95\% CL upper limit on BR($B_s \to \mu^+ \mu^-$).
The continuous (black) lines are contours of $\tan \beta$.
}
\label{fig:EMO}
\end{figure}

Fig.~\ref{fig:EMO} displays a portion of the no-scale SU(5) parameter 
space for $\lambda^\prime = 2$ and $ \lambda = - 0.1$ that is compatible with these
new constraints. There is a region at low $M_{in}$ that is excluded because the LSP
would be charged (shaded brown), a region at large $m_{1/2}$ and high $M_{in}$
where we do not find consistent solutions of the RGEs (shaded ochre), and a region
at low $m_{1/2}$ that is excluded by $b \to s \gamma$ (shaded green). The ratio of supersymmetric Higgs
vevs, $\tan \beta$, is determined dynamically, with the indicated values along the continuous
black contours, and values of $m_H$ calculated with {\tt FeynHiggs}~\cite{FH} are
shown as dash-dotted (red) contours. 

The area with $m_{1/2} \lsim 840$~GeV (purple solid
line) is excluded by ATLAS searches for supersymmetric particles at the LHC~\cite{ATLASsusy}.
There are (dark blue) strips of parameter space extending up to $(m_{1/2}, M_{in}) \sim (1300, 10^{17.5})$~GeV and with $m_{1/2} \sim 1500$ GeV and $M_{in} \sim 10^{18}$~GeV where the relic cold dark matter density is consistent with 
cosmology and $m_H \sim 125$~GeV. In view of the
estimated theoretical uncertainty of $\sim 3$~GeV in the {\tt FeynHiggs} calculation, all the portions of
these dark matter strips with $m_{1/2} \gsim 1000$~GeV
may be consistent with the LHC measurement of $m_H$. This region is also compatible with
the experimental measurement of $B_s \to \mu^+ \mu^-$: the diagonal (green) solid line marks
the 95\% CL upper limit on this decay, including theoretical uncertainties~\cite{bmm}. Regions with suitable $m_H$ 
and satisfying the other constraints can also be found
for somewhat smaller values of $\lambda^\prime$ and $|\lambda|$. 

An alternative to the pure no-scale approach taken above
is the possibility that the terms in the K\"ahler potential
corresponding to the Standard Model fields are contained either only
partly within the no-scale logarithm (\ref{nsKahler}), or outside it entirely
via minimal kinetic terms $K \ni |y_{SM}|^2$. 
For example, in the latter case, and assuming a superpotential of the 
form (\ref{WSM}), the universal soft scalar mass term is now $m_0 = r = m_{3/2}$
whilst the relations for $A_0$ and $B_0$ are still given by (\ref{mBA}).
It is interesting to note that for the case $\alpha = \beta = 0$, one
recovers the pure gravity-mediated models discussed in \cite{dlmmo,pgm}. Viable 
phenomenological models can be constructed with input universality input above the GUT 
scale~\cite{dlmmo} as in the models discussed above, or with GUT-scale universality 
at relatively low $\tan \beta \simeq 2$ \cite{pgm}. Fixing $\tan \beta$ and still satisfying the
supergravity boundary condtions for $A_0$ and $B_0$ can be 
achieved with the addition of a Giudice-Masiero term in the K\"ahler potential \cite{GM}.

Our analysis demonstrates that no-scale supergravity provides a suitable framework
for addressing many problems in cosmology and particle physics, including inflation,
leptogenesis, neutrino masses and dark matter as well as the LHC measurement of
$m_H$ and limits on supersymmetric particles. Concerning future experimental tests,
we note that future CMB experiments may probe the predictions of Starobinsky-like
inflationary models such as ours, and that higher-energy LHC running will be sensitive
to $m_{1/2} \lsim 1500$~GeV, the range suggested in the sample models displayed
in Fig.~\ref{fig:EMO}. Thus there are both cosmological and collider prospects for
exploring the ideas presented here.

\noindent {\bf Acknowledgements. }  We thank Djuna Croon, T. Gherghetta,  
J. Evans, Thomas Hahn, Sven Heinemeyer,
Nick Mavromatos and Georg Weiglein for discussions.
The work of J.E. was supported in part by
the London Centre for Terauniverse Studies (LCTS), using funding from
the European Research Council 
via the Advanced Investigator Grant 267352.
The work of D.V.N. was supported in part by the
DOE grant DE-FG03-95-Er-40917.
The work of K.A.O. was supported in part
by DOE grant DE--FG02--94ER--40823 at the University of Minnesota.

\end{document}